\documentclass[10pt,aps,prl,reprint]{revtex4-2}

\usepackage{bm}
\usepackage{graphicx}
\usepackage{appendix}
\usepackage{amssymb,amsfonts,amsmath}
\usepackage{adjustbox}

\usepackage{booktabs}
\usepackage{silence}
\usepackage{nameref}

\usepackage[bookmarks = false, pdfstartview = {FitH}, colorlinks = true, allcolors = blue]{hyperref}

\makeatletter
\renewcommand{\@fnsymbol}[1]{\ensuremath{\ifcase#1\or \dagger\or *\or \ddagger\or
   \mathsection\or \mathparagraph\or \|\or **\or \dagger\dagger
   \or \ddagger\ddagger \else\@ctrerr\fi}}
\makeatother

\def\Emory{\vspace{+0.4ex}Department of Physics, Emory University, Atlanta, GA 30322\vspace{+0.4ex}}

\begin{document}

\title{A microstructural rheological model for transient creep in polycrystalline ice}

\author{Alex J. Vargas}
\thanks{These authors contributed equally to this work.}
\affiliation{\Emory}

\author{Ranjiangshang Ran}
\thanks{These authors contributed equally to this work.}
\affiliation{\Emory}

\author{Justin C. Burton}
\thanks{Corresponding author: justin.c.burton@emory.edu\vspace{+0.4ex}}
\affiliation{\Emory}

\date{\today}

\begin{abstract}
The slow creep of glacial ice plays a key role in sea-level rise, yet its transient deformation remains poorly understood. Glen’s flow law, where strain rate is simply a function of stress, cannot predict the time-dependent creep behavior observed in experiments. Here we present a physics-based rheological model that captures all three regimes of transient creep in polycrystalline ice. The key components of the model are a series of Kelvin-Voigt mechanical elements that produce a power-law (Andrade) creep, and a single viscous element with microstructure and stress dependence that represents reorientation in the polycrystalline grains. 
The interplay between these components produces a minimum in the strain rate at approximately 1\% strain, which is a universal but unexplained feature reported in experiments. Due to its transient nature, the model exhibits fractional power-law exponents in the stress dependence of the strain rate minimum, which has been conventionally interpreted as independent physical processes. Taken together, we provide a compact, mechanistic framework for transient ice rheology that generalizes to other polycrystalline materials and can be integrated into constitutive laws for ice-sheet models.
\end{abstract}

\maketitle 


Accurately predicting future sea-level rise hinges on reliable models of polar ice sheet dynamics, which depend sensitively on the rheology of ice \cite{greve2009dynamics,vanderveen2013fundamentals}. Polar regions are warming nearly four times faster than the global average \cite{Rantanen2022,IPCC2021}, accelerating the mass loss of the Greenland and Antarctic ice sheets that together store over 99\% of Earth’s freshwater ice \cite{Shepherd,Otosaka2023}. 
Under high emissions scenarios, global sea-level rise by 2100 could be as high as 1 meter \cite{Siahaan2022,siegert2020twenty, van2022high}. Projections of this kind are controlled by large-scale ice sheet models, whose flow laws are rooted in constitutive descriptions of ice rheology that are time-independent, yet much of the deformation in glaciers and ice streams occurs under transient conditions \cite{cuffey2010physics,pattynF,Alley_Bentley_1988,vaughan}. In particular, transient creep at constant stress, relevant for slowly evolving stress fields in grounded ice, provides a window into the time-dependent mechanical response of polycrystalline ice. A physics-based understanding of this behavior is essential for modeling features like internal stratigraphy, ice stream onset zones, and memory effects in flow history \cite{gillet2006flow,CHRISTIANSON201457,Raymond_1983}.

At the core of ice sheet modeling lies the constitutive law governing ice flow under stress, with most models adopting Glen’s empirical flow law \cite{glen1955creep}:
\begin{equation}
    \dot{\varepsilon}=A\sigma^n,
\end{equation}
where $\dot\varepsilon$ is the effective strain rate, $\sigma$ is the effective stress, $n$ is the stress exponent, and $A$ is a prefactor that depends on temperature and can have fractional units depending on the value of $n$.
However, this framework lacks any explicit time dependence and cannot capture the transient, history-sensitive deformation observed in both laboratory and field settings \cite{ashby1985creep,budd1989review,journaux2019recrystallization}. Notably, Glen originally formulated his law by fitting the stress dependence of the minimum strain rate in creep experiments under constant stress. The minimum, $\dot\varepsilon_\text{min}$, occurs near 1\% strain in isotropic polycrystalline ice, and is robust to the presence of impurities \cite{song2005creep,hammonds2018effects}. This choice implicitly folds transient dynamics into a time-independent flow law, obscuring the role of internal evolution in shaping long-term flow behavior. 

Despite these known limitations, refinements to Glen’s flow law, most notably the composite flow law proposed by Goldsby and Kohlstedt \cite{goldsby2001superplastic,goldsby1997grain,goldsby1997flow}, remain fundamentally anchored in steady-state assumptions. Their framework combines several deformation mechanisms (e.g., diffusion creep, grain-boundary sliding, basal slip, and dislocation creep) by summing individual strain rates, but it is built entirely from strain-rate minima measured in constant-stress experiments. As a result, it neglects the transient evolution that precedes and follows the minimum and offers no mechanism for microstructure evolution or structural memory, and remains the dominant framework for modeling of glacier flow and interpreting remote sensing data \cite{fan2025flow,ranganathan2024modified,wang2025deep}. Moreover, the deformation mechanisms each have different non-integer stress exponents (i.e., $n = 1.8$ or $n = 2.4$) that originate from empircal fitting, violate dimensional consistency without auxiliary stress scales, and further obscure physical interpretation. Thus, these models are inherently limited in their ability to capture the rich, time-dependent creep behavior documented in laboratory studies and natural ice cores.

To address this, we develop a physics-based rheological model that captures both transient creep and microstructural evolution. The model consists of a series of Kelvin–Voigt mechanical elements that generate power-law (Andrade) creep \cite{Andrade1910}, combined with a single viscous element whose microstructure- and stress-dependent behavior captures reorientation within the polycrystalline grains. When combined, these two components produce a strain rate minimum consistently occurring at $\approx 1$\% strain and apparent non-integer exponents in experimentally-relevant yet transient regimes. The model reproduces experimental creep data with physically meaningful parameters and exponents, providing a compact framework that links microstructural evolution to macroscopic flow, and can be directly incorporated into large-scale ice-sheet models.




\begin{figure}
    \centering
    \includegraphics[width=1\linewidth]{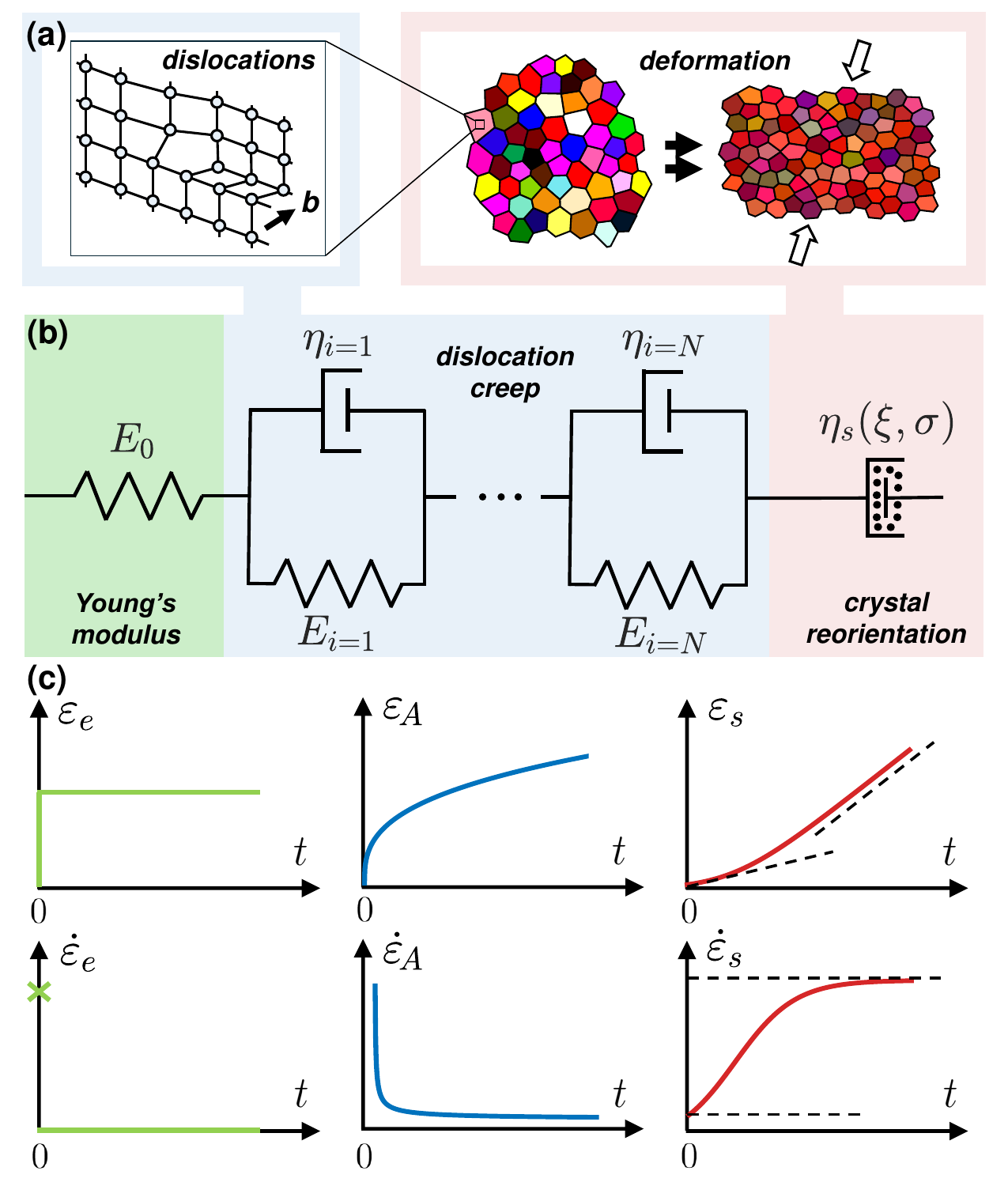}
    \caption{Rheological model of polycrystalline ice. (a) Microscopic origin of ice deformation. Within individual grains, deformation occurs by the motion of dislocations. Simultaneously, this motion gradually aligns the grains’ $c$-axes, shown as increasingly uniform grain colors.
    (b) The mechanical analog of the model consists of three components: an elastic spring (green) with modulus $E_0$, a series of Kelvin–Voigt elements (blue) producing power-law (Andrade) creep, and a structural dashpot (red) with viscosity $\eta_s$ dependent on microstructure $\xi$ and stress $\sigma$.
    (c) Schematics of strains, $\varepsilon$, and strain rates, $\dot\varepsilon$, as functions of time under a step change in stress, for each component in the mechanical analog.}
    
    \label{fig:model_circuit}
\end{figure} 

The deformation of polycrystalline ice is governed by dislocation activity within individual grains and by the evolving crystal orientation and grain size distribution, collectively referred to as the ice “fabric” \cite{pimienta1987mechanical,duval1995dynamic}.
Under constant stress, ice exhibits a characteristic three-stage creep behavior: an initial power-law decay in strain rate, a secondary nonmonotonic minimum, and a tertiary regime approaching steady flow \cite{budd1989review,cuffey2010physics}.
These stages reflect the interplay between irreversible dislocation avalanches along the basal planes \cite{weertman1983creep,duval1983rate,weiss1997acoustic,weiss2004dislocation} and long-timescale microstructural reorganization such as grain rotation and recrystallization \cite{duval1995dynamic,gillet2006flow}.
To capture these features within a unified, physically motivated framework, we propose a composite yet minimal rheological model that reproduces the universal behavior observed in transient ice creep experiments.

Figure~\ref{fig:model_circuit}(a)--(b) illustrates the mechanical analog of our model, which is composed of three components connected in series. The instantaneous elastic response of ice is represented by a spring with modulus $E_0$ (green), followed by two nonlinear elements that emulate distinct physical mechanisms. The initial power-law creep observed in ice is modeled by an infinite series of Kelvin-Voigt elements (blue) with retardation times that follow a power-law distribution. The infinite series respresents generalized Andrade creep behavior, where the strain evolves according to $\varepsilon(t) \propto t^{1/p}$. We interpret this behavior as dislocation avalanches---intermittent, burst-like rearrangements that unjam the internal structure, similar to behavior observed in glassy and granular materials \cite{miguel2002dislocation,louchet2009andrade,richeton2005breakdown}.

The last element [Fig.~\ref{fig:model_circuit}(b), red] in the series is a structural dashpot with a viscosity, $\eta_s(\xi, \sigma)$, that evolves with the ice fabric. Here, the microstructural parameter, $\xi(t) \in [0,1]$, quantifies the degree of misalignment in the grain orientation distribution, with $\xi = 1$ representing an isotropic fabric and $\xi = 0$ the steady state orientation. It evolves through the ordinary differential equation, $d\xi/dt = -\xi(t)/\tau_s$, progressively decreasing the viscosity from a hard state $\eta_1=\eta_s(\xi = 1)$ to a soft state $\eta_{0}=\eta_s(\xi = 0)$. We borrow the concept of microstructural evolution from thixotropic models of structured fluids \cite{ran2023understanding,larson2019review,wei2018multimode}.
Under sustained stress, individual grains rotate their $c$-axes toward orientations favorable for basal slip \cite{van1994development,rigsby1960crystal,azuma1985formation}, whose dynamics is governed by kinematics analogous to Jeffery’s equation \cite{gillet2005user,gillet2006flow,supp}. This reorientation process defines the stress dependence of the structural timescale: $\tau_s(\sigma)=(k \sigma^p)^{-1}$ \cite{supp}. Since both Andrade creep and fabric evolution are facilitated by the motion of dislocations, we have chosen the same stress exponent $p$ for both terms. This choice  also enables the model to reproduce the experimentally observed strain-rate minimum near 1\% strain.


The total strain is written as the sum of the strains of the elastic spring, $\varepsilon_e(t)$, the Andrade components, $\varepsilon_A(t)$, and the structural dashpot, $\varepsilon_s(t)$:
\begin{equation}
    \varepsilon(t) = \varepsilon_e(t) + \varepsilon_A(t) + \varepsilon_s(t),
\end{equation}
where the individual strains and strain rates are illustrated in Fig.~\ref{fig:model_circuit}(c).
Each Kelvin-Voigt element in the Andrade component has an elastic modulus $E_i$ and a viscosity $\eta_i$ (see End Matter). The strain of each Kelvin-Voigt element, $\varepsilon_i(t)$, evolves as:
\begin{equation}
    \dot{\varepsilon}_i(t) + \frac{E_i}{\eta_i}\,\varepsilon_i(t)
    = \frac{1}{\eta_i}\,\sigma(t).
\end{equation}
The strain rate of the structural dashpot is given by:
\begin{equation}
  \dot{\varepsilon}_s(t)=\frac{\sigma(t)}{\eta_{s}(\xi,\sigma)},
  \qquad
  \eta_s(\xi,\sigma)=\eta_0+\xi(t)\Delta\eta,
  \label{Eq:structural}
\end{equation}
where $\Delta\eta=\eta_{1} - \eta_{0}$. Mechanically, this is equivalent to two dashpots ($\eta_0$ and $\Delta\eta$) connected in parallel. 
Both viscosities are made shear-thinning with $\eta_{0}=\sigma^{1-n_0}/A_0$ and $\Delta\eta=\sigma^{1-n_1}/A_1$, where $A_0$ and $A_1$ are temperature-dependent prefactors that follow an Arrhenius relation \cite{vanderveen1990flow,bird1987dynamics}, and $n_0$ and $n_1$ are power-law exponents of the shear-thinning behavior. 

\begin{figure}[t!]
    \centering
    \includegraphics[width=1\linewidth]{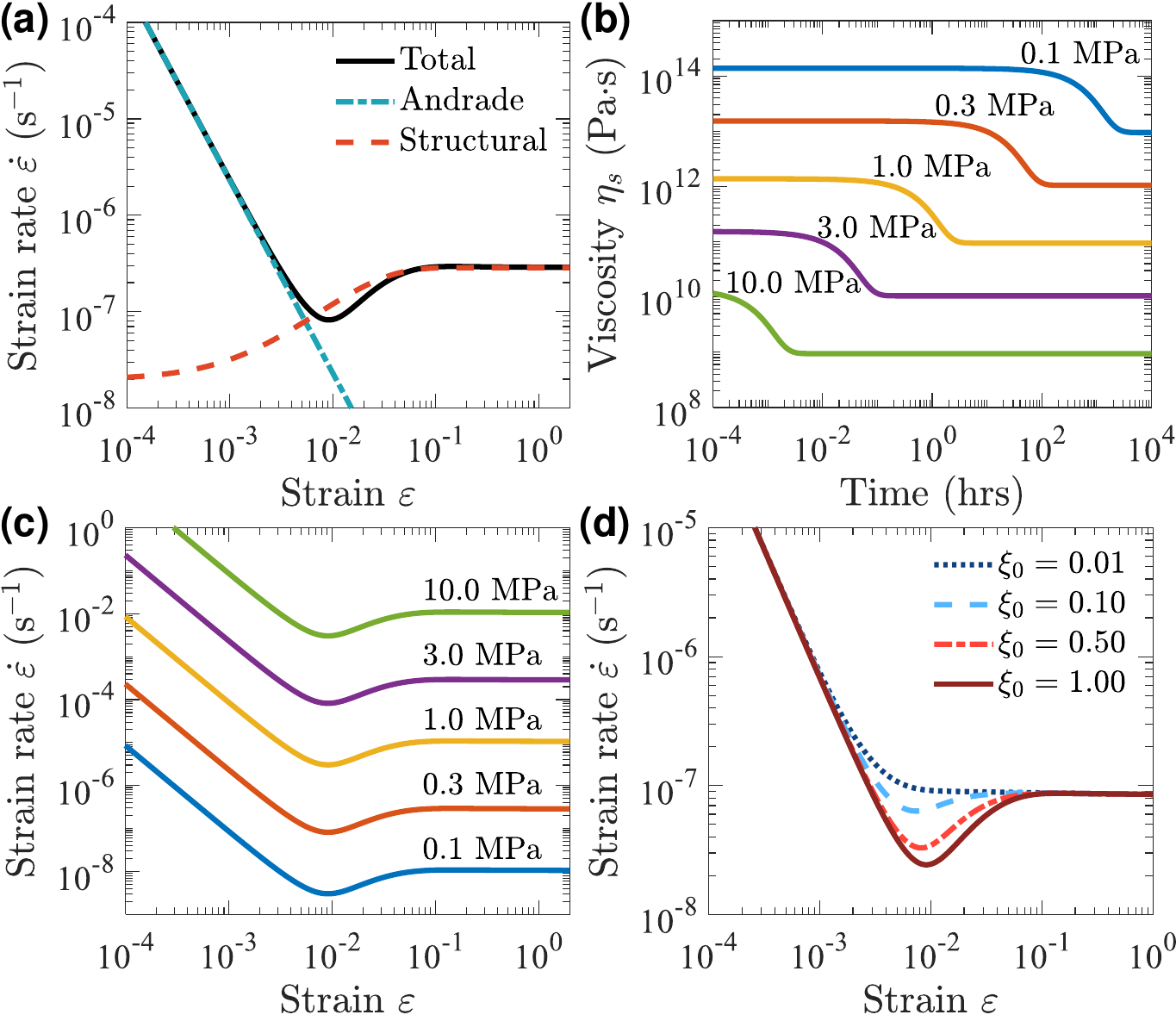}
    \caption{Mechanical response of the model under constant stress conditions. (a) Total strain rate (black) decomposed into Andrade (blue) and structural dashpot (red) contributions. Primary creep is Andrade-dominated, secondary reflects crossover, and tertiary is controlled by structural evolution. (b–c) Stress dependence of effective viscosity and strain rate, demonstraing softening in the tertiary regime and a minimium consistently at $\approx1$\%, respectively. (d) Effect of the initial microstructural parameter, $\xi_0$, which captures anisotropic versus isotropic initial conditions observed in experiments. For all curves, the stress exponents were $p=3$, $n_0=3$, and $n_1=3$.}
    \label{fig:model_power}
\end{figure}

The model yields an analytical solution under constant stress, the condition of laboratory creep tests (see End Matter):
\begin{align}
    \varepsilon(t>0) &= \frac{\sigma}{E_0}
    + \beta \sigma t^{1/p}
    + \frac{\sigma \tau_s}{\eta_{0}}\,
      \ln\!\left(\frac{e^{t/\tau_s} + \lambda}{1 + \lambda}\right),
      \label{eq:strain} \\
    \dot{\varepsilon}(t>0) &= \frac{\beta}{p}\,\sigma t^{(1-p)/p}
    + \frac{\sigma}{\eta_{0}\!\left(1 + \lambda e^{-t/\tau_s}\right)}.
      \label{eq:strainrate}
\end{align}
%
%
%
Here, $\lambda = \xi_{0}\Delta\eta/\eta_{0}$ is a dimensionless parameter that combines the ratio of isotropic and anisotropic viscosities with the initial degree of microstructural disorder, $\xi_{0}=\xi(t=0)$. Together, Eqs.~\eqref{eq:strain} and \eqref{eq:strainrate} capture the full transient creep response of polycrystalline ice under constant stress conditions. 
Importantly, this model predicts stress-dependent softening and effective viscosity evolution without requiring additional fitting parameters beyond those tied to physical processes.

\begin{figure}[t!]
    \centering
    \includegraphics[width=2.8in]{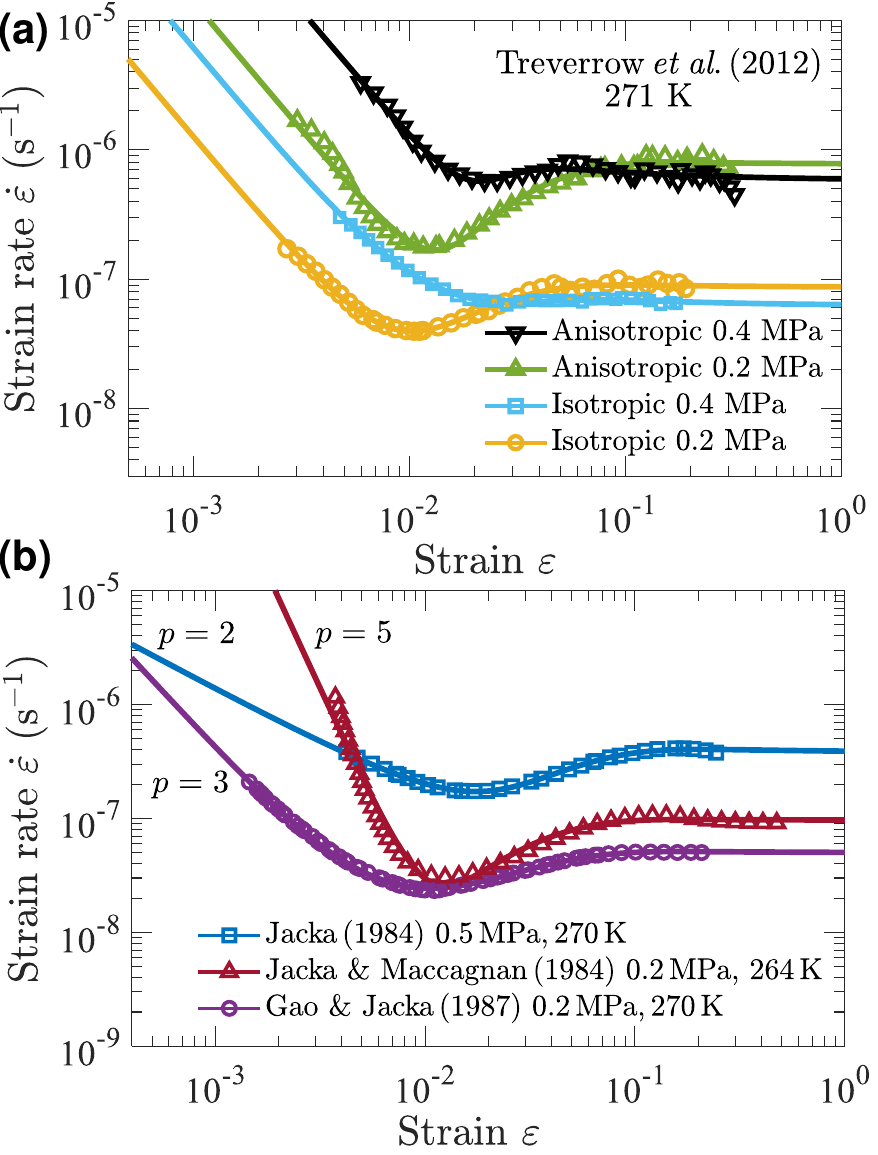}
    \caption{Model fits to experimental data using Eqs.~\ref{eq:strain} and \ref{eq:strainrate}. Markers are experimental data; solid curves are model fits. (a) Data from Treverrow \textit{et al.} \cite{Treverrow2012tertiary} showing differences in creep behavior between anisotropic and isotropic ice samples. The stress exponent used for the fits is $p=3$. (b) Data from references \cite{jacka1984laboratory,jackaMacc1984ice,gao1987approach} showing that $2\lesssim p\lesssim 5$, depending on experimental conditions. All fitting parameters are listed in Table S1-S4.}
    \label{fig:fit_across_data}
\end{figure}

Figure~\ref{fig:model_power} illustrates the distinct stages of ice deformation and their dependence on internal structure and applied stress in the model. Figure~\ref{fig:model_power}(a) decomposes the total strain rate into Andrade and structural components, revealing a clear crossover. Early-time deformation is dominated by dislocation-mediated creep, while long-time behavior reflects the progressive softening of the internal microstructure. This transition produces a transient, nonmonotonic strain-rate minimum. Figure~\ref{fig:model_power}(b) shows the time-dependent softening of the effective viscosity ($\eta_s$) associated with the microstructural evolution. Figure~\ref{fig:model_power}(c) shows that our model reproduces a key experimental trend: the strain rate miniminum appears near 1\% strain, independent of stress \cite{budd1989review}. Finally, Fig.~\ref{fig:model_power}(d) shows the influence of initial fabric anisotropy, encoded by $\xi_0$, on the transient response. The strain rate in anisotropic samples decreases monotonically because $\xi_0\approx 0$ and $\eta_s\approx\eta_0$ throughout the creep process (Eq.~\ref{Eq:structural}). Isotropic samples display a nonmonotonic strain rate because $\eta_s$ evolves over the timescale $\tau_s$ as the $c$-axes of the ice grains align. Together, these results underscore how the interplay of dislocations and microstructure produces transient flow behavior consistent with constant creep experiments of polycrystalline ice.

Using the Markov chain Monte Carlo (MCMC) method \cite{ran2023understanding,ran2025electrostatics,supp,foreman2013emcee}, we directly fit our model in Eqs.~\eqref{eq:strain} and \eqref{eq:strainrate} to data from creep experiments. Figure~\ref{fig:fit_across_data}(a) shows data from Treverrow \textit{et al.}~\cite{Treverrow2012tertiary} for isotropic and anisotropic polycrystalline ice samples, where the experiments last for $\sim 4$ months. The model captures the complete creep response with the minimum consistently occurring at $\approx1$\% strain in isotropic samples, independent of stress level. In anisotropic samples, the $c$-axes are nearly aligned at the beginning of the experiment, thus no such minimum appears. The structural viscosity already approaches the tertiary value of $\eta_0$, corresponding to $\lambda\approx 0$. Figure~\ref{fig:fit_across_data}(b) shows that the power-law creep exponent $p$ can depend on stress, temperature, and sample preparation. Our model fits indicate that $p$ ranges from $2$ to $5$, which is consistent with reported creep exponents for ice and a broad range of polycrystalline materials \cite{duval2010creep,louchet2009andrade,nechad2005creep,nabarro2006creep,gribb1998low,zhang2021investigation}. 

\begin{figure}
    \centering
    \includegraphics[width=1\linewidth]{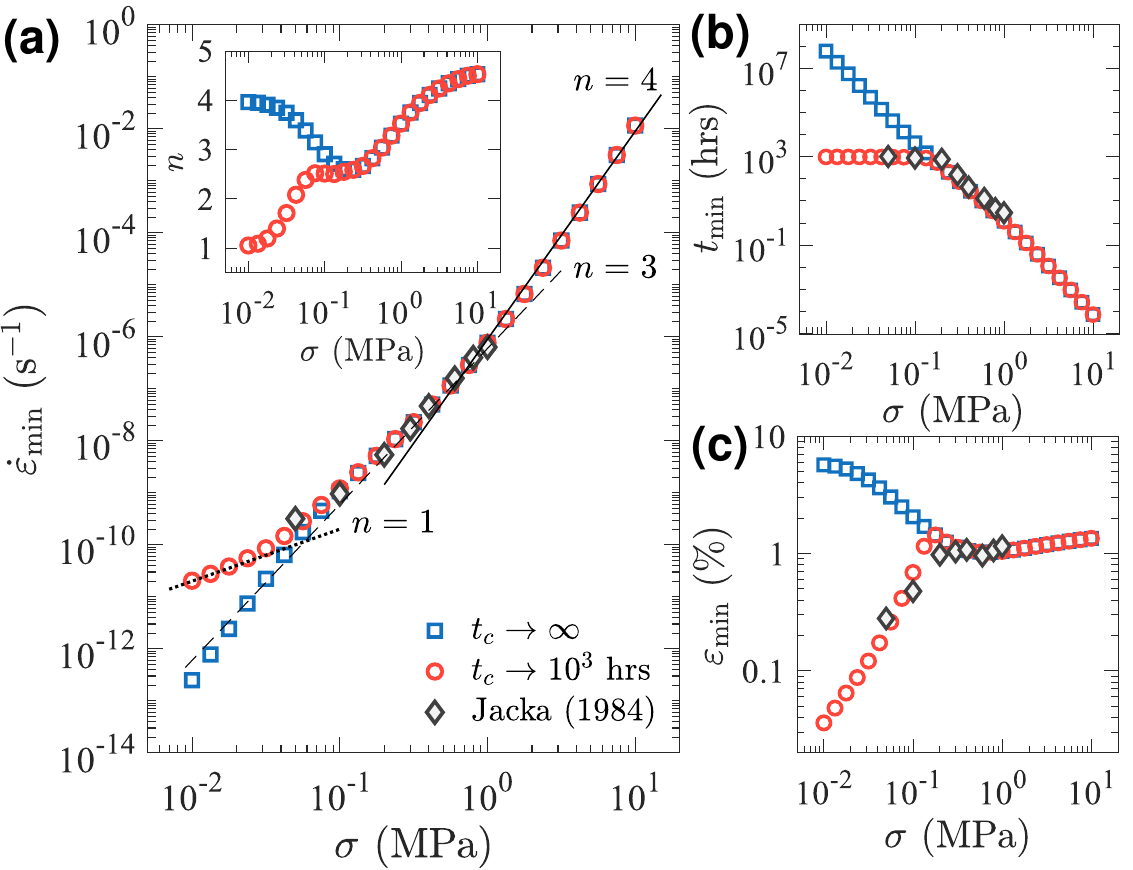}
    \caption{
    Different flow regimes can emerge with apparent power-law exponents in transient dynamics.
    (a) Minimum strain rate vs. stress (blue squares) using $p=5$, $n_0=4$, and $n_1=2$, demonstrating that different power-law exponents can be observed from a single model with transient dynamics. The inset indicates the apparent exponent, which is the derivative in $\log$ space. The red circles show $\dot\varepsilon_\text{min}$ at a fixed time, $t_c$, which emulates limited experimental observation times. The black diamonds are data from \citet{jacka1984time}. 
    (b) The time, $t_{\min}$, to reach $\dot\varepsilon_\text{min}$ decreases with stress, yet plateaus at $t_c$. (c) The strain at the minimum, $\varepsilon_\text{min}$, is nearly constant at $\approx1$\%, but can deviate at low stresses when $p\neq n_0$, or with finite $t_c$.
    }
    \label{fig:minima_scaling}
\end{figure}

While the fits validate the model against laboratory data, they also highlight deeper implications for how ice rheology is traditionally interpreted. Figure~\ref{fig:minima_scaling}(a) illustrates why the long-standing practice of identifying the strain-rate minima with a steady-state flow law, central to Glen’s flow law \cite{glen1955creep,cuffey2010physics,vanderveen2013fundamentals} and the Goldsby and Kohlstedt composite flow relation \cite{goldsby2001superplastic,fan2025flow,wang2025deep,ranganathan2024modified}, is problematic. Their framework represented a major advance in systematically collating laboratory creep data, but it relied on the assumption that the minima define unique material constants. Our model shows instead that the apparent stress dependence at the minima is emergent: instantaneous slopes ranging from $n\approx 2.4$ to 4 arise from the crossover between Andrade-type transient creep and microstructural softening. Interpreting such values as evidence for multiple deformation mechanisms conflates a dynamical crossover with intrinsic steady-state rheology, creating the appearance of distinct flow laws where none exist.

A further complication is that reported data do not always reach the strain rate minimum due to a limited observation window. 
Lower stresses push the strain rate minimum to progressively later times, often well beyond the weeks to months accessible in laboratory tests. In Fig.~\ref{fig:minima_scaling}(a), this manifests as larger strain rates and a smaller effective stress exponent at low stress (orange data). Figure~\ref{fig:minima_scaling}(b) shows that the time to reach the strain rate minimum, $t_\text{min}$, plateaus at low stress due to experimental limitations. Additionally, at low stress when $p\neq n_0$, the strain rate minimum can deviate from $\approx1$\%, as illustrated in Fig.~\ref{fig:minima_scaling}(c). This conflation of data from different regimes is problematic for steady-state flow laws. For example, \citet{goldsby2001superplastic} use data taken at $\varepsilon\approx1$\% at higher stresses, and $\varepsilon\lesssim0.1$\% at lower stresses \cite{butkovich1960creep}.
Thus, power law relationships between $\dot\varepsilon_\text{min}$ and $\sigma$ may not represent fundamental steady states but time- and stress-dependent crossovers in behavior. This underscores the need for models that explicitly resolve both transient and microstructural dynamics, as we do here.



In summary, we have shown that the transient creep behavior of polycrystalline ice can be understood through a minimal and physically motivated model that combines Andrade-type dislocation activity with a microstructure-dependent viscous response. The combination of these processes naturally produces the primary power-law creep, the secondary creep with a strain rate minimum near $\varepsilon\approx 1$\%, and the transition to tertiary flow. 
In our model, the apparent non-integer stress exponents reported in classical analyses emerge as transient signatures of evolving microstructure rather than indicators of different deformation mechanisms. Instead, the exponents $p$, $n_0$, and $n_1$ are constrained by distinct physical processes (e.g., $p$ is determined by dislocation jamming \cite{miguel2002dislocation}). 

Further progress would benefit from renewed constant stress experiments that extend fully into the tertiary regime and include associated fabric measurements. In addition to quantitatively interpreting laboratory creep data over a wide range of experimental conditions, our framework provides a dynamical constitutive law for transient ice rheology that can be employed in ice sheet modeling. We expect this to be particularly important in fast flowing regimes, or regions where the strain rate changes rapidly, such as near grounding lines of ice shelves and tidewater glaciers. Furthermore, this general framework can be applied to other polycrystalline materials such as olivine \cite{maor2025learning,hansen2014protracted,wallis2017dislocation,silber2025plastic,korenaga2008new}, which exhibit comparable microstructural controls on transient creep.

\section*{Acknowledgments}
This work was supported by the Keck Foundation and the Gordon and Betty Moore Foundation, Grant DOI: 10.37807/gbmf12256. R.R. also acknowledges funding from the Tarbutton Postdoctoral Fellowship of Emory College of Arts and Sciences.

\section*{Data Availability} 

The experimental datasets used for fitting are digitized from Treverrow \textit{et al.}~\cite{Treverrow2012tertiary}, \citet{gao1987approach}, \citet{jacka1984laboratory}, and \citet{jackaMacc1984ice}. 

\bibliographystyle{apsrev4-2}
\bibliography{references.bib}

\section*{End Matter}

\textbf{Power-law creep (statistical approach)}---We can use a series of $N$ Kelvin-Voigt elements to represent power-law (Andrade) creep [Fig.~\ref{fig:model_circuit}(b)], with the elastic modulus of each element fixed as: $E_i=N\!E_1$. Under a constant stress $\sigma$, each Kelvin-Voigt element obeys the following dynamics:
\begin{align}
    \tau_i&=\eta_i/N\!E_1,\\
    \varepsilon_i&=\dfrac{\sigma}{N\!E_1}\left(1-e^{-t/\tau_i}\right),\\
\varepsilon&=\sum\limits_{i=1}^{N}\varepsilon_i.
\end{align}
Here, $\tau_i$ is the intrinsic timescale associated with the $i$-th element, and each element contributes a strain $\sigma/N\!E_1$ in the limit $t\rightarrow\infty$. Thus the total strain is given by $\varepsilon$. For an infinite number of Kelvin-Voigt elements, we convert the sum to an integral over a distribution of elements with different timescales:
\begin{equation}
\varepsilon=\sum\limits_{i=1}^{\infty}\varepsilon_i=\dfrac{\sigma}{E_1}\int_{\tau_\text{min}}^{\tau_\text{max}}\left(1-e^{-t/\tau}\right)P(\tau)d\tau\label{fint},
\end{equation}
where $P(\tau)$ is a probability density function (PDF), and $\tau_\text{min}$ ($\tau_\text{max}$) is the minimum (maximum) timescale in the distribution. In our model, $\varepsilon\propto t^{1/p}$, where $p=3$ is typically associated with Andrade creep. This behavior is gauranteed by the normalized PDF:
\begin{gather}
P(\tau)=\dfrac{1}{p}\dfrac{\tau^{(1-p)/p}}{ \left(\tau_\text{max}^{1/p}-\tau_\text{min}^{1/p}\right)},\label{pfunc}\\
\int_{\tau_\text{min}}^{\tau_\text{max}}P(\tau)d\tau=1.
\end{gather}
Note that $P(\tau)$ follows a power-law distribution whose scaling exponent is set by $p$: smaller $p$ (e.g., $p = 3$) leads to a heavier-tailed distribution with greater weight at long timescales, whereas larger $p$ (e.g., $p = 5$) yields a lighter-tailed distribution with larger weight at shorter timescales.
Plugging Eq.~\ref{pfunc} into Eq.~\ref{fint} and integrating yields the total strain:
\begin{equation}
\varepsilon=\dfrac{\sigma}{E_1}\left[1+\dfrac{t^{1/p}}{p}\dfrac{\Gamma\left(-\dfrac{1}{p},\dfrac{t}{\tau_\text{min}},\dfrac{t}{\tau_\text{max}}\right)}{\tau_\text{max}^{1/p}-\tau_\text{min}^{1/p}}\right],
\end{equation}
where $\Gamma(\alpha,s_1,s_2)$ is the generalized incomplete gamma function:
\begin{equation}
\Gamma(\alpha,s_1,s_2)=\int_{s_1}^{s_2}x^{\alpha-1}e^{-x}dx.
\end{equation}
We associate the maximum and minimum timescales with the collective motion of dislocations in the material. For example, a short timescale can be due to a few localized dislocations, while a long timescale can represent a delocalized, cooperative motion of many dislocations. To obtain a power law, we first take the limit that $\tau_\text{min}\rightarrow 0$ and $\tau_\text{max}\rightarrow \infty$, where the leading order term is:
\begin{equation}
\varepsilon=\Gamma\left(1-\frac{1}{p}\right)\dfrac{\sigma}{E_1}\left(\dfrac{t}{\tau_\text{max}}\right)^{1/p}.
\end{equation}
Comparing this result with the power-law term in Eq.~\ref{eq:strain}, we find that:
\begin{equation}
\beta\propto\dfrac{1}{E_1\tau_\text{max}^{1/p}}.
\end{equation}
The constant $\beta$ can be fixed if we simultaneously make $E_1$ smaller as $\tau_\text{max}\rightarrow\infty$, which results in a larger maximum strain, $\sigma/E_1$. This ensures that the strain will grow without bound for power-law creep. 

\textbf{Power-law creep (spectrum approach)}---In the statistical approach, the elastic modulus for each Kelvin-Viogt element is fixed as a constant, $E_i = N\!E_1$, and the number density of elements with the timescale $\tau_i$ is described by the PDF, $P(\tau)$. Alternatively, one can assign a fixed number of Kelvin-Viogt elements associated with the timescale $\tau_i$, while varying their elastic modulus $E_i$ according to a spectrum, $E(\tau)$, to reproduce the same power-law creep behavior. This corresponds to a Prony-series representation \cite{christensen1982theory}: 
\begin{equation}\label{Prony}
    \varepsilon=\sum\limits_{i=1}^{N}\dfrac{\sigma}{E_i}\left(1-e^{-t/\tau_i}\right).
\end{equation}
For each Kelvin-Viogt element, we assign a distinct timescale $\tau_i$, which is constructed to be uniformly and logarithmically spaced in $[\tau_{\min},\tau_{\max}]$:
\begin{equation}
\tau_i=\tau_{\min} r^{i-1},\qquad \tau_{\max}=\tau_N=\tau_{\min} r^{N-1},
\end{equation}
where $r=\tau_{i+1}/\tau_i>1$ is a ratio. Under the construction and in the limits of $N\to\infty$ and $r\to1$, the sum in Eq. \eqref{Prony} converges to the integral \cite{supp}:
\begin{equation}\label{Prony cont}
    \varepsilon=\sigma\int_{\tau_{\min }}^{\tau_{\max }} \frac{1}{\tau E(\tau)}\left(1-e^{-t / \tau}\right) d\tau.
\end{equation}
Comparing Eq. \eqref{Prony cont} with Eq. \eqref{fint}, one would notice that $E(\tau)\propto [\tau P(\tau)]^{-1}$. In the limits of $\tau_\text{min}\to 0$ and $\tau_\text{max}\to \infty$, it can be shown that the strain in Eq. \eqref{Prony cont} converges to the desired power-law form, $\epsilon=\beta\sigma t^{1/p}$, given the following $E(\tau)$ spectrum \cite{supp}:
\begin{equation}
    E(\tau) = \Gamma\!\left(1-\frac{1}{p}\right)\frac{p}{\beta\tau^{1/p}}.
\end{equation}

\textbf{Strain at the strain rate minimum}---Our model assumes the same exponent $p$ in both power-law creep [Eq.~\eqref{eq:strain}] and the microstrucural timescale, $\tau_s=(k\sigma^{p})^{-1}$. This leads to a total strain at the strain rate minimum which is invariant to stress, as observed in experiments. The strain and strain rate from the Andrade, power-law creep is:
\begin{align}
    \varepsilon&=\beta\sigma t^{1/p},\\
    \dot\varepsilon&=\dfrac{1}{p}\beta\sigma t^{(1-p)/p}.
\end{align}
Solving the first equation for $t$, and plugging into the second equation, we see that
\begin{equation}
\dot\varepsilon=\dfrac{1}{p}\left(\beta\sigma\right)^{p}\varepsilon^{1-p}.
\end{equation}
Thus, the timescale associated with Andrade creep is $\sim 1/(\beta\sigma)^p$, which is the same stress scaling used for $\tau_s$. Since $\dot\varepsilon_\text{min}$ results from two distinct processes with independent timescales, the position of the minimum is roughly invariant to changes in stress if both timescales scale with stress in the same way. 

\textbf{Model with time-varying stress}---Our analysis of the model shown in Fig.\ref{fig:model_circuit}(b) was performed under constant stress, where the strain for each element can be computed independently, and then summed. In the case that $\sigma=\sigma(t)$, we can express the strain use Green's functions. Consider the solution of a single Kelvin-Voigt element to a stress impulse:
\begin{align}
    \delta(t-t')&=E_1(G_1+\tau_i\dot G_1),\\
    G_1(t-t')&=\dfrac{1}{E_1\tau_i}e^{-(t-t')/\tau_i}\theta(t-t')\label{green1},
\end{align}
where $\theta(x)$ is the Heaviside step function. While $G_1$ is the solution of a single Kelvin-Voigt element, we can compute the Green's function for the infinte series as well:
\begin{gather}
G(t-t')=\int_{\tau_\text{min}}^{\tau_\text{max}}G_1(t-t')P(\tau)d\tau\\
=\dfrac{(t-t')^{1/p-1}\Gamma\left(\dfrac{p-1}{p},\dfrac{t-t'}{\tau_\text{max}},\dfrac{t-t'}{\tau_\text{min}}\right)}{pE_1\left(\tau_\text{max}^{1/p}-\tau_\text{min}^{1/p}\right)}\theta(t-t').
\end{gather}
For a known, time-dependent stress, $\sigma(t)$, the power-law component of our model can then be expressed as:
\begin{equation}
\varepsilon=\int_{-\infty}^{\infty}\sigma(t')G(t-t')dt'
\end{equation}

\end{document}